# Giant Strengthening of Superconducting Pairing in Metallic Nanoclusters.

Yurii N. Ovchinnikov[a,b] and Vladimir Z. Kresin[c8]

[a)] Max-Planck Institute for Physics of Complex Systems, Dresden,

D-01187, Germany

[b)] L.D.Landau Institute for Theoretical Physics, Russian Academy of

Sciences, 117334, Moscow, Russia

[c)] Lawrence Berkeley Laboratory, University of California at Berkeley,

CA 94720

Abstract

The presence of shell structure and the accompanying high level degeneracy leads to a strengthening of the pairing interaction in some metallic nanoclusters. It is predicted that for specific systems one can expect a large increase in the values of the critical temperature and other parameters.

This paper contains a theoretical analysis of the superconducting state of metallic nanoclusters. We show that under special, but perfectly realistic, conditions such clusters should display a drastic increase in the values of superconducting parameters such as the critical temperature and the energy gap. The superconducting state of nanoparticles is a subject of many interesting experimental and theoretical studies (see, e.g., [1-10] and the review [11]). If metallic clusters contain a small number of electrons (N≈ $10^2$-$10^3$), one might think that they do not display superconducting properties, because the average level spacing ($E_F$/N~$10^2$-$10^3$ meV) greatly exceeds the pairing energy gap. However, the situation is more complicated. The fact of the matter is that there exist clusters in which the pattern of electronic states is very different from that of a simple equally spaced level distribution. They contain highly degenerate electronic levels, or groups of very close levels (quasi-degenerate case). The importance of the so-called shell structure for the superconducting state was indicated in [5,10] and, especially in [ 6 ]. Below, we focus precisely on such a situation.

As is known, metallic clusters contain delocalized electrons whose states organize into shells, similar to those in atoms or nuclei [12]. In some clusters, shells are completely filled all the way up to the highest occupied shell (HOS): e.g., those with N=$N_m$=20,40,58,92,138,168,.... These are known as "magic" numbers, see, e.g., the reviews [13,14] (N denotes the number of delocalized electrons and $N_m$ specific "magic" numbers). Such clusters are spherical. The electronic states in such magic-number clusters are labeled by their orbital momentum $l$ and radial quantum number $n$, and if $l$ is large, the shell is highly degenerate. In addition, the energy spacing ΔE between neighboring shells varies, and some of them are separated by only a small ΔE. One can show (see below) that the combination of high degeneracy and a small energy spacing between the HOS and the lowest unoccupied shell (LUS) leads to the possibility of a large strengthening of superconducting pairing in the corresponding clusters.

Let us write down the main equations for the pairing order parameter. We employ the thermodynamic Green's function formalism allowing one to evaluate the order parameter $\Delta(\omega_n)$ and the critical temperature $T_c$ ( here $\omega_n=(2n+1)\pi T$, see, e.g., [15]). Since we expect some clusters to display a high $T_c$, so that the ratio $2\pi T_c/\tilde{\Omega}$ (where $\tilde{\Omega}$ is the characteristic phonon frequency) will not be very small, we do not restrict ourselves to the usual BCS weak coupling approximation (corresponding to $2\pi T_c \ll \tilde{\Omega}$) and consider instead the more general equation (cf.[16]-[18]). For systems with a discrete energy spectrum this equation reads

$$\Delta(\omega_n)Z = \frac{\eta T}{V} \sum_{n',j} D(\omega_n - \omega_{n'}) g_j \Delta(\omega_{n'}) \left[\omega^2_{n'} + \Delta^2(\omega_{n'}) + (E_j - \mu)^2\right]^{-1} \qquad (1)$$

Here $D = \tilde{\Omega}^2[(\omega_n-\omega_{n'})^2+\tilde{\Omega}^2]^{-1}$ is the phonon Green's function, $\Delta(\omega_n)\left[\omega_n^2 + \Delta^2(\omega_n) + (E_i - \mu)^2\right]^{-1}$ is the Gor'kov's function, V is the cluster volume, $\tilde{\Omega}$ is the characteristic phonon frequency, $g_j = 2(2l+1)$ is the degeneracy and $\varepsilon_j-\mu$ is the electron energy referred to the chemical potential. Z is the renormalization function; here we shall not write out its explicit expression (see, e.g., [17,18]). As is known, the presence of Z removes the divergence at $\omega_n=\omega_{n'}$. The parameter $\eta$ describes the electron-phonon coupling and has the following form: $\eta=<I^2>/(M\tilde{\Omega}^2)$; here M is the ionic mass, I is the electon-ion interaction, and $<I^2>$ is the matrix element averaged over the states involved in the pairing [19,20]. Note that the value of $\eta$ is close to its bulk value $\eta_b$. Indeed, the surface of the cluster can be treated as a scatterer (cf. [21]) and therefore the pairing is analogous to that in the case of a "dirty" superconductor analyzed in [3], see also [22], whereby the mean free path is much shorter than the coherence length. Then the average value of $I^2$ is not affected by the scattering and, indeed, $\eta \approx \eta_b$ where $\eta_b$ is the Hopfield parameter (see,e.g.,[20]). Note also that the characteristic vibrational frequency $\tilde{\Omega}$ is close to the bulk value because the pairing is mediated mainly by the short-wavelength part of the vibrational spectrum. Strictly speaking, the order parameter depends on j. However, for shells close to the HOS (when $k_H R \gg 1$, i.e., when $N > 10^2$) this dependence is rather weak (see, e.g., [10,23]) and can be neglected. Here R is the cluster radius and $k_H$ is the electron wave vector at HOS ($k_H \approx 2/r_s$, where $r_s$ is the electron density parameter). The value of $k_H$ for the clusters of interest is close to the Thomas-Fermi screening wave vector and to the value of the Fermi momentum $k_F$ ( we put $\hbar = 1$). The Fermi energy $E_F$ is likewise close to $E_H$, the energy of the HOS. As the cluster size increases, a continuum energy spectrum develops, and we obtain the equation [24].

In order to calculate the value of $T_c$, one puts $\Delta=0$ in the denominator on the right-hand side of Eq.(1), then $\omega_n=(2n+1)\pi T_c$. Incorporating the explicit expression for Z, we obtain after some manipulations the following matrix equation (cf. [17,18]):

$$\Delta(\omega_n) = \sum_{n'} K_{nn'} \Delta(\omega_{n'}) \quad ; \quad n,n' \geq 0 \quad (2)$$

where

$$K_{nn'} = (\lambda T \Omega^2 / 2\nu_b V)(f^+_{nn'} + f^-_{nn'} - 4\delta_{nn'}\omega_n^2 f^+_{nn'} f^-_{nn'}) \sum_j g_j [\omega^2_{n'} + (E_j - \mu)^2]^{-1}$$

Here $f_{nn'}^{+(-)} = [(\omega_n \pm \omega_{n'})^2 + \tilde{\Omega}^2]^{-1}$. We introduce the dimensionless parameter $\lambda = \eta\nu_b$, where $\nu_b = mk_F/2\pi^2$. Since $\eta \approx \eta_b$ and $E_h \approx E_F$, the parameter $\lambda \approx \lambda_b$, the bulk coupling constant (see, e.g., [25]) which corresponds to the bulk critical temperature $T_c^b$; the values for $\lambda$ are known for many superconductors (see, e.g., [26]). Therefore, Eqs. (2,3) allow us to evaluate $T_c$ by using known parameters. We can therefore focus directly on the impact of size quantization, especially the degeneracy caused by the shell structure.

For some spherical clusters (see below) the degeneracy $g_j$ is quite large. This is a very important factor for our analysis, since it plays the role of an effective increase in the value of $\lambda$. It is also essential that the HOS-LUS interval (analogous to the "HOMO-LUMO" spacing in molecular spectroscopy) is not large. Then the term $\xi_j = \varepsilon_j - \mu$ in the denominator is relatively small (see below), which is also an important factor.

Since the number of electrons N is fixed, we also have an equation determining the position of the chemical potential:

$$N = \sum_j g_j \{1 + \exp[(E_j - \mu)/T]\}^{-1} \quad (4)$$

For a closed-shell cluster the chemical potential lies in the middle of the HOS-LUS interval. At finite temperatures $\mu \equiv \mu(T)$ shifts in accordance with Eq.(4). It is convenient to write the expression for $\mu$ in the form

$$\mu = E_H + \tilde{\mu} E_H[(E_L/E_H) - 1] \qquad (5)$$

Here $E_{H(L)} \equiv E_{HOS(LOS)}$. Within the potential box model the ratio $E_L/E_H$ is $E_L/E_H = (Z_L/Z_H)^2$, where $Z_L$ and $Z_H$ are the roots of the Bessel function $J_{l+1/2}(x)$ corresponding to the neighboring terms $E_L$ and $E_H$.

Based on Eqs. (2-4), we can calculate $T_c$. Let us apply these general equations to a specific case, for example a cluster with N=168. The reason for such a choice will be seen below. Note, first of all, that this is a closed-shell cluster (see, e.g., [27]). The cluster with N=168 contains fully occupied shells up to the HOS with $l=7$. As a result, the degeneracy $g_j = 2(2l+1)$ is very high: $g_H=30(!)$. Moreover, the next shell is relatively close, so that $(E_L/E_H)-1 \approx 5 \cdot 10^{-3}$. Note that the LUS also has a large degeneracy: $g_L=18$.

The critical temperature can be calculated from Eq. (2), or more specifically, from the equation:

$$\text{Det}|1-K_{nn'}| = 0, \qquad (6)$$

where the matrix $K_{nn'}$ is defined by Eq.(3).

To be even more specific, let us first look at $In_{56}$. Indeed, indium is a bulk superconductor with $T_c=3.4K$ and so it is interesting to consider the impact of size quantization on its properties. It has been observed experimentally (see, e.g., [28]) that In clusters display shell structure. Since In, like Al, has three weakly bound electrons, the $In_{56}$ cluster indeed contains 168 electrons.

Let us evaluate the value of $T_c$ for the In clusters with N=168. The following parameter values are used for In: $\lambda \approx 0.55$, $\tilde{\Omega} \approx 8meV$ [26], $k_F \approx 1.5 \cdot 10^8 cm^{-1}$, $R \approx 7 \cdot 10^{-8} cm$ (see the discussions following Eqs.(1) and (3)). The value of $\lambda$ is modified to account for the Coulomb pseudopotential (see, e.g., [18]).

With the use of the $In_{56}$ parameters, one can solve the eigenvalue problem for Eq. (2), that is, solve Eq. (6). As indicated above, the quantities $\xi_j = \varepsilon_j - \mu$ are not large for j=H,L (H and L correspond to HOS and LUS, respectively); these two terms make the major contribution to the sum in Eq. (6). For accuracy, we also have included the terms next to H and L. All other terms could be neglected. It turns out that the solutions rapidly converges (cf. [17,18]), and it is sufficient to solve Eq. (6) as a 2x2 matrix.

The calculation yields $T_c \approx 21.5$ K (!). This greatly exceeds the aforementioned bulk value $T_c^b = 3.4$ K. Such a large increase in $T_c$ is caused by the high degeneracy of HOS and LUS and by the small magnitude of $E_{H(L)} - \mu$. Note that the solution is self-consistent, and the use of the matrix equation (6) with fast convergence is justified.

It is essential that the value of Tc of a nanocluster is not universal but depends strongly on its parameters. A remarkably high value of Tc is predicted for $Nb_{168}$ clusters. The following parameters are employed: $\lambda \approx 1.05$, $E_F \approx 5.3$ eV, $\tilde{\Omega} \approx 1.5 \cdot 10^{-2}$ eV, $R \approx 9$A, $E_0 \equiv E_{HOS} \approx 6.45$ eV. With the use of Eqs.(3-6) we obtain $T_c \approx 10^2$ K (!). This value greatly exceeds that for bulk Nb, $T_c^b = 9.2$ K. A similarly high value can be obtained for Zn clusters.

Let us discuss the case of clusters with incomplete shells. These are different from the "magic" clusters in two important aspects. Clusters with partially unoccupied shells undergo a Jahn-Teller shape distortion (see, e.g., [14]). This splits the degenerate level, which is not favorable for pairing. On the other hand, a removal of electrons from the HOS strongly affects the position of the chemical potential. For example, at T=0K it now coincides with the highest occupied level. This factor can enhance Tc. The best scenario would correspond to nanoclusters with slightly incomplete shells (e.,g., with N=166) and with an appropriate vibrational force matrix (elastic constants), such that the shape deviations from sphericity would be relatively weak. Note once again that the effect is not universal and strongly depends on the parameters of the material It turns out (the detailed analysis will be described elsewhere) that such situation occurs for Zn clusters with N=166. For such clusters the value of $T_c$ is of order of $T_c \approx 120$ K (!).

According to a very interesting paper [6], a spherical cluster with a half-filled shell should have a high value of $T_c$. However, in this situation the shape deformation would be very large, drastically decreasing the value of $T_c$. The author [6] suggested that it might be possible to use a cluster network combined with charge transfer to overcome this problem. We think that the situation with almost filled-shell looks promising, because the deformation could be rather small.

The role of fluctuations grows with decreasing particle size (see, e.g., [29]). However, for the clusters studied here the Ginzburg parameter $(\Delta/E_F)^2$ is still relatively small, although the impact of fluctuation should be taken into account e.g., in the study of a.c. properties.

Pair correlation can manifest itself via an increased magnitude of the HOS-LUS interval (revealing the presence of the energy gap); it is important to note that the magnitude of this interval strongly depends on the temperature. The pairing should manifest itself also in odd-even effects for cluster spectra and in their magnetic properties. The phenomenon is also promising for the creation of high Tc tunneling networks. These aspects will be discussed in detail elsewhere.

In summary, small metallic nanoclusters which possess a large degeneracy $2(2l+1)$ of the highest occupied shell and a small HOS-LUS interval are predicted to display a giant strengthening of the superconducting pair correlation. A similar effect also occurs for the clusters with slightly incomplete shells and, correspondingly, small shape deformation.

The authors are grateful to J.Friedel, A.Goldman, V.V.Kresin and M.Tinkham for fruitful discussions.

The research of YNO was supported by the CRDF under Contract No. RP1-2565-MO-03 and by RFBR (Russia).

*) Contacting author: vzkresin@lbl.gov